\documentstyle[preprint,aps]{revtex}
\begin{document}
\draft

\title{Scaling Model of Annihilation-Diffusion Kinetics for
Charged Particles with Long-Range Interactions}
\author{Sergei F. Burlatsky$\dagger$,
 Valeriy V. Ginzburg and Noel A. Clark}
\address{$\dagger$Department of Chemistry, University of Washington, Seattle WA 98195-1700\\
 Department of Physics, University of Colorado, 
 Boulder CO 80309-0390}
\date{}
\maketitle
\begin{abstract}We propose the general scaling model for the diffusion-annihilation reaction
$A_{+} + A_{-} \longrightarrow \emptyset$ with long-range power-law interactions. The presented
scaling arguments lead to the finding of three different regimes, depending on the space dimensionality d and the long-range force power exponent n. The obtained kinetic phase diagram agrees well
with existing simulation data and approximate theoretical results.
\end{abstract}


The problem of annihilation-diffusion kinetics has generated significant interest in
recent years, since it was shown~\cite{Burlatsky} - \cite{Lebowitz} that in a bimolecular reaction
$A + B \longrightarrow \emptyset$, long-wavelength density fluctuations cause significant slowing down
of the particle density decay. While the ordinary kinetic-rate approach suggests that in a bimolecular
reaction, particle density $\rho$ always decays as $t^{-1}$, the account of fluctuations leads to a 
power-law decay:

\begin{equation}
	\rho \propto t^{- \nu},
	\label{eq:genscaling}
\end{equation}

where $\nu \leq 1$; when no long-range interaction exists between particles, $\nu = d/4$ for $d < 4$ and
$\nu = 1$ when $d \geq 4$, where d is space dimensionality~\cite{Burlatsky} - \cite{Zeldovich}. Thus,
4 is critical dimension for the bimolecular reaction, and below it, long-wavelength fluctuations determine
annihilation kinetics entirely.

Even though the role of fluctuations in systems without long-range interactions is well understood and 
confirmed by numerical simulations~\cite{Wilczek} - \cite{Leyvraz}, the situation is far less clear when 
long-range forces are present. Meanwhile, the latter problem is especially important in different physical
applications, e.g., in condensed matter physics, when one replaces particles $A$ and $B$ with dislocations
or vortices and analyzes their annihilation upon quench from a high-temperature to a low-temperature phase. These considerations prompted several numerical studies into the problem of annihilation kinetics in
Coulombic systems in two dimensions~\cite{Jang}, \cite{DeGrand}, \cite{Mondello}. While it was shown that power-law decay~(\ref{eq:genscaling}) is satisfied on a late stage of the annihilation process, and that exponent $\nu$ is
rather close to 1, its value has not been determined fully or calculated rigorously. In our recent work
~\cite{Jang}, numerical simulations in two dimensions yielded $\nu = 0.9 \pm 0.05$. The scaling theory was
proposed to explain this unusual result~\cite{Jang},\cite{Ginzburg}, based on an assumption that the system
first relaxes charge density fluctuations towards statistical equlibrium and only then proceeds with
annihilation. However, other theories~\cite{BOO1} - \cite{Krapivsky},  predicted $\nu = 1$ for Coulombic systems, if initial conditions remain random when annihilation starts. 

Quite recently, we proposed a self-consistent description of the annihilation-diffusion kinetics~\cite{Ginzburg96}, which allows to take into account different initial conditions. This theory neglects fluctuations of the total particle density, but, unlike a mean-field theory, takes into account the fluctuations of the charge density. Such an approximation enables one to solve equations of evolution and determine asymptotics for different scaling regimes. In the case of Gaussian random initial charge density, kinetic phase diagram was calculated and final (large-t) asymptotics were as follows:

if $n \geq 1 + d/2$ and $d < 4$ (fluctuation-dominated region)

\begin{equation}
	\rho \propto (Dt)^{- \nu}, \hspace{1in} \nu = d/4;
\label{eq:bzotw}
\end{equation}

if $n < 1 + d/2$ and $n \geq d - 1$ (intermediate region)

\begin{equation}
	\rho \propto (Qt)^{- \nu}, \hspace{1in} \nu = \frac{d}{2 - d + 2n};
\label{eq:intermed}
\end{equation}

and if $d \geq 4$ (mean-field region)

\begin{equation}
	\rho \propto ({\cal K} t)^{- \nu}, \hspace{1in} \nu = 1,
\label{eq:meanfield}
\end{equation}

with D, Q and ${\cal K}$ being diffusion, electrostatic and annihilation constants. Equations~(\ref{eq:bzotw}) - (\ref{eq:meanfield}) do not exhaust all scaling solutions of self-consistent equations of evolution, these are only the large-t asymptotics. Equations~(\ref{eq:bzotw}) and ~(\ref{eq:meanfield}) are very well-known, while the intermediate solution~(\ref{eq:intermed}) was proposed in the scaling theory of Ispolatov and Krapivsky~\cite{Krapivsky} (but the region of its applicability was determined in a different manner).

In this Letter, we propose a simple scaling theory to derive relations~(\ref{eq:bzotw}) - (\ref{eq:meanfield}). Such a theory would not only support the validity of the self-consistent approach of Ref.~\cite{Ginzburg96}, but also would give an additional insight on the role of initial conditions in determining late stage asymptotics for annihilation-diffusion problems.

Let us start by writing a Langevin equation for the i-th particle, assuming that inertia forces are negligible in comparison with viscous drag force:

\begin{equation}
	\beta \frac{d {\bf r_{i}}}{dt} = \sum_{j} \frac{Q q_{i} q_{j}}{|{\bf r_{i} - r_{j}}|^{n+1}}
	({\bf r_{i} - r_{j}}) + {\bf f_{ri}},
\label{eq:langevin}
\end{equation}

where $\beta$ is a "friction coefficient", and ${\bf f_{ri}}$ is a random Brownian force acting on the i-th particle. We will use the Langevin equation to characterize a slow relaxation of an initial particle density fluctuation and thus estimate the annihilation rate.

Let us select a region with characteristic size L. Because of the assumption about randomness of the initial particle distribution, $\delta N \propto \sqrt{N}$, where N is the number of particles inside the volume $V=L^{d}$. The mean-square normalized density fluctuation in this volume is:

\begin{equation}
	\frac{\delta \rho}{\rho_{0}} \propto L^{-d/2}.
\label{eq:msfluct}
\end{equation}

As it is traditionally done in scaling theories of annihilation, we assume that excess particles start annihilating only when they travel the distance on the order of L. In this case, assuming that diffusion is irrelevant, we rewrite the Langevin equation, replacing all occurences of ${\bf r}$ with L:

\begin{equation}
	\beta \frac{d L}{dt} = \frac{\delta Q(L)}{L^{n}},
\label{eq:sclang}
\end{equation}

where total charge fluctuation in the region $\delta Q(L) \propto L^{d/2}$, again due to the Gaussian initial conditions. From this follows:

\begin{equation}
	L^{n+1-d/2} \propto t,
\label{eq:tvsl}
\end{equation}

or

\begin{equation}
	L \propto t_{L}^{\frac{1}{n+1-d/2}},
\label{eq:lvst}
\end{equation}

where $t_{L}$ represents time necessary for an average particle to travel the distance L (assuming it
does not annihilate during this time).  At the time $t_{L}$, majority of particles that originally were in the volume V, have annihilated, and only those excess particles still remained, so the density is:

\begin{equation}
	\frac{\rho(t_{L})}{\rho_{0}} \approx \frac{\delta N}{N} \propto L^{-d/2} 
	\propto t_{L}^{- \frac{d}{2(n+1-d/2)}},
\label{eq:intscal1}
\end{equation}

or, omitting the index "L",

\begin{equation}
	\rho \propto t^{- \frac{d}{2n + 2 - d}}.
\label{eq:intscal2}
\end{equation}

In order for scaling laws (\ref{eq:tvsl}) and (\ref{eq:intscal2}) to be valid, it is necessary that i) drift occurs slower than annihilation, and that ii) drift occurs faster than thermal (Brownian) diffusion. These conditions imply that:

\begin{equation}
	n + 1 - d/2 \leq 2,
\label{eq:difcond}
\end{equation}

\begin{equation}
	\frac{d}{2n + 2 - d} \leq 1,
\label{eq:anncond}
\end{equation}

which is equivalent to: $d - 1 \leq n \leq 1 + d/2$. 

If condition (\ref{eq:difcond}) is not satisfied, i.e., $n > 1 + d/2$, deterministic drift is irrelevant, and everything is determined by the competition of diffusion and annihilation. In this case, as expected:

\begin{equation}
	L \propto t^{1/2},
\label{eq:diflength}
\end{equation}

\begin{equation}
	\rho \propto t^{-d/4}, \hspace{0.5in} d \leq 4;
\label{eq:bzoagain}
\end{equation}

\begin{equation}
	\rho \propto t^{-1}, \hspace{0.5in} d > 4.
\label{eq:mfagain}
\end{equation}

Thus, using simple scaling approximations, we confirmed validity of the self-consistent approximation for the annihilation-diffusion system with long-range interactions. Scaling exponents obtained in the large-t limit agree well with known theoretical models and numerical simulations. We also showed that the initial density distribution plays an important role in determining large-t scaling behavior at least for strongly interacting systems such as Coulombic, and made estimates of this behavior when the initial distribution is Gaussian.

In the future, more theoretical and numerical efforts will be required to explore in more details the rich kinetic phase diagram for systems with long-range interactions. While the use of scaling arguments like the one proposed here cannot be used for exact analysis of annihilation behavior, it can and will complement other numerical and analytical methods in elucidating main features and dependencies of annihilation kinetics.

{\bf Acknowledgement}. This work was done during the March Meeting of the American Physical Society (St. Louis, March 1996).

\end{document}